\documentclass{article}

\usepackage{microtype}
\usepackage{graphicx}
\usepackage{booktabs} 
\usepackage{xspace}
\usepackage{array}
\usepackage{multirow}
\usepackage{makecell}
\usepackage{amsmath}
\usepackage{xcolor}
\usepackage{graphicx}
\usepackage{subcaption}

\usepackage{hyperref}

\newcommand{\sys}{\textsf{Bolt}\xspace}

\usepackage[accepted]{mlsys2021}

\mlsystitlerunning{Bolt: Bridging the Gap between Auto-tuners and Hardware-native Performance}

\newcommand{\leyuan}[1]{{{\color{red}(Leyuan) #1}}}

\begin{document}

\twocolumn[





\mlsystitle{Bolt: Bridging the Gap between Auto-tuners \\ and Hardware-native Performance}







\mlsyssetsymbol{equal}{*}

\begin{mlsysauthorlist}
\mlsysauthor{Jiarong Xing}{equal,bd,rice}
\mlsysauthor{Leyuan Wang}{equal,bd}
\mlsysauthor{Shang Zhang}{nv}
\mlsysauthor{Jack Chen}{nv}
\mlsysauthor{Ang Chen}{rice}
\mlsysauthor{Yibo Zhu}{bd}
\end{mlsysauthorlist}

\mlsysaffiliation{bd}{ByteDance} 
\mlsysaffiliation{nv}{NVIDIA}
\mlsysaffiliation{rice}{Rice University}

\mlsyscorrespondingauthor{Leyuan Wang}{leyuan.wang@bytedance.com}


\vskip 0.3in

\begin{abstract}
Today's auto-tuners (e.g., AutoTVM, Ansor) generate efficient tensor programs by navigating a large search space to identify effective implementations, but they do so with opaque hardware details. Thus, their performance could fall behind that of hardware-native libraries (e.g., cuBLAS, cuDNN), which are hand-optimized by device vendors to extract high performance. On the other hand, these vendor libraries have a fixed set of supported functions and lack the customization and automation support afforded by auto-tuners. \sys is based on the recent trend that vendor libraries are increasingly modularized and reconfigurable via declarative control (e.g., CUTLASS). It enables a novel approach that bridges this gap and achieves the best of both worlds, via \textit{hardware-native templated search}. \sys provides new opportunities to rethink end-to-end tensor optimizations at the graph, operator, and model levels. \sys demonstrates this concept by prototyping on a popular auto-tuner in TVM and a class of  widely-used platforms (i.e., NVIDIA GPUs)---both in large deployment in our production environment. \sys improves the inference speed of common convolutional neural networks by 2.5x on average over the state of the art, and it auto-tunes these models within 20 minutes. 
\end{abstract}
]




\printAffiliationsAndNotice{\mlsysEqualContribution} 

\section{Introduction}
\label{sec:intro} 

Auto-tuning engines~\cite{autoTVM, ansor, halide-auto-scheduler, flextensor} are at the heart of a variety of DNN compilers and frameworks~\cite{tvm, xla, tensorflow, pytorch}. Example auto-tuners like AutoTVM~\cite{autoTVM} and Ansor~\cite{ansor} infer hardware cost models from afar, by executing sample implementations on a particular device and observing their performance. Building on the inferred cost models, auto-tuners take tensor programs as inputs, and navigates a large search space to select effective transformations for high performance. Operating with opaque device models affords generality, as the same approach can be applied to different hardware, without requiring hardware details.


As a downside, treating devices as opaque models comes with performance implications---for any specific device, it is likely that hardware-native performance as delivered by low-level vendor libraries is hard to come by. Traditional vendor libraries (e.g., cuBLAS~\cite{cublas}, cuDNN~\cite{cudnn}) expose a fixed set of primitives that are heavily hand-optimized for the underlying hardware. For workloads that fit into their library primitives, and for users with expert device knowledge, directly using such libraries extracts hardware-native performance. For instance, auto-tuners like AutoTVM and Ansor do not achieve competitive performance compared to cuBLAS and cuDNN for non-FP32 compute-intensive operators on NVIDIA GPUs because of their inefficient usage of tensor cores. In our benchmarks, Ansor~\cite{ansor} only achieves  20\% of cuBLAS performance for FP16 GEMMs on NVIDIA Tesla T4 GPUs (see Figure~\ref{fig:vscublas} for more details). 



Related, opaque device models also lead to a prolonged auto-tuning time, 
as the search process is less informed by hardware details. 
For instance, it takes AutoTVM~\cite{autoTVM} 10 hours on x86 CPUs and 7 days on NVIDIA GPUs to tune 
all workloads in the ResNet-50 model~\cite{lorien}. This has led to the development 
of special databases~\cite{tophub} that cache and reuse tuning logs, but this approach 
only goes so far. Models have increasing dynamism, not only in terms of dynamic 
data structures~\cite{liang2016semantic} but also dynamic shapes~\cite{bert}, 
making caching much less effective. Maintaining these databases also incurs substantial costs. 

Can we achieve the best of both worlds, combining the flexibility of auto-tuners \textit{and} the hardware-native performance as afforded by vendor implementations? \sys bridges this gap leveraging the trend that vendor libraries are increasingly \textit{templated}, reconfigurable with declarative parameters to suit different scenarios, but exposing concise interfaces that are potentially amenable to  auto-tuning. 
An exemplar of a templated design, NVIDIA CUTLASS~\cite{cutlass}, encodes efficient design patterns but is not committed to a fixed set of primitives. Users can parameterize the templates to suit their workloads, and they can extend and compose templated primitives for new functionality. In a similar vein, Intel OneDNN~\cite{onednn} and AMD ROCm~\cite{rocm} also share this emerging trend for their platforms.  

We propose to enable end-to-end DNN model optimizations via \textit{hardware-native templated search} utilizing the above trend.  \sys operates on hardware-native templates which interposes a thin layer between the high-level computational graph and the low-level CUDA code, opening up opportunities for joint optimization. First, it generates tensor programs with hardware-native performance via efficient templated search. Second, by composing and extending template primitives, \sys enables novel computational graph-level optimizations. Combined, \sys enables auto-tuners to achieve both \textit{graph-level} and \textit{operator-level} optimization and generates the implementations with hardware-native performance using a significantly shortened turnaround time. Furthermore, \sys also enables \textit{model-level} optimizations by proposing system-model codesign principles. If models are designed in a system-friendly manner, they can fully utilize the benefits of \sys and achieves more efficient inference.

We prototype \sys in TVM~\cite{tvm} for NVIDIA GPUs utilized its open-sourced CUTLASS library, while noting that the new design approach generalizes beyond this scenario. Compared to Ansor, \sys achieves 2.5x inference speedup on widely-used convolutional neural networks; it auto-tunes these workloads within 20 minutes. Our new computational graph level optimization---persistent kernel fusion---leads to a performance improvement up to 1.3x and 2.0x on GEMMs and Conv2Ds. Finally, we validate our model-level optimization---system-model codesign---by augmenting RepVGG models~\cite{repvgg}, which effectively improves model accuracy with less speed sacrifice.
\sys is deployed in our production setting, serving real models and workloads, and will be released in open source to the community.

\section{Background and motivation}
\label{sec:background}

\subsection{Auto-tuners have a performance gap} 


State-of-the-art DNN compilers and frameworks~\cite{autoTVM, ansor} leverage auto-tuning to identify effective tensor implementations. Auto-tuners transform a tensor program into an equivalent but structurally different one, which delivers higher performance on the target. This is achieved by constructing a cost model of the hardware via learning---e.g., by building a training set with sample programs and their performance on the target, and by predicting which implementations are likely to be performant when navigating the search space. Operating afar from the hardware delivers benefits such as platform generality, but it also leads to two performance implications. 



\begin{figure}[t!]
\centering\includegraphics[width=0.9\linewidth]{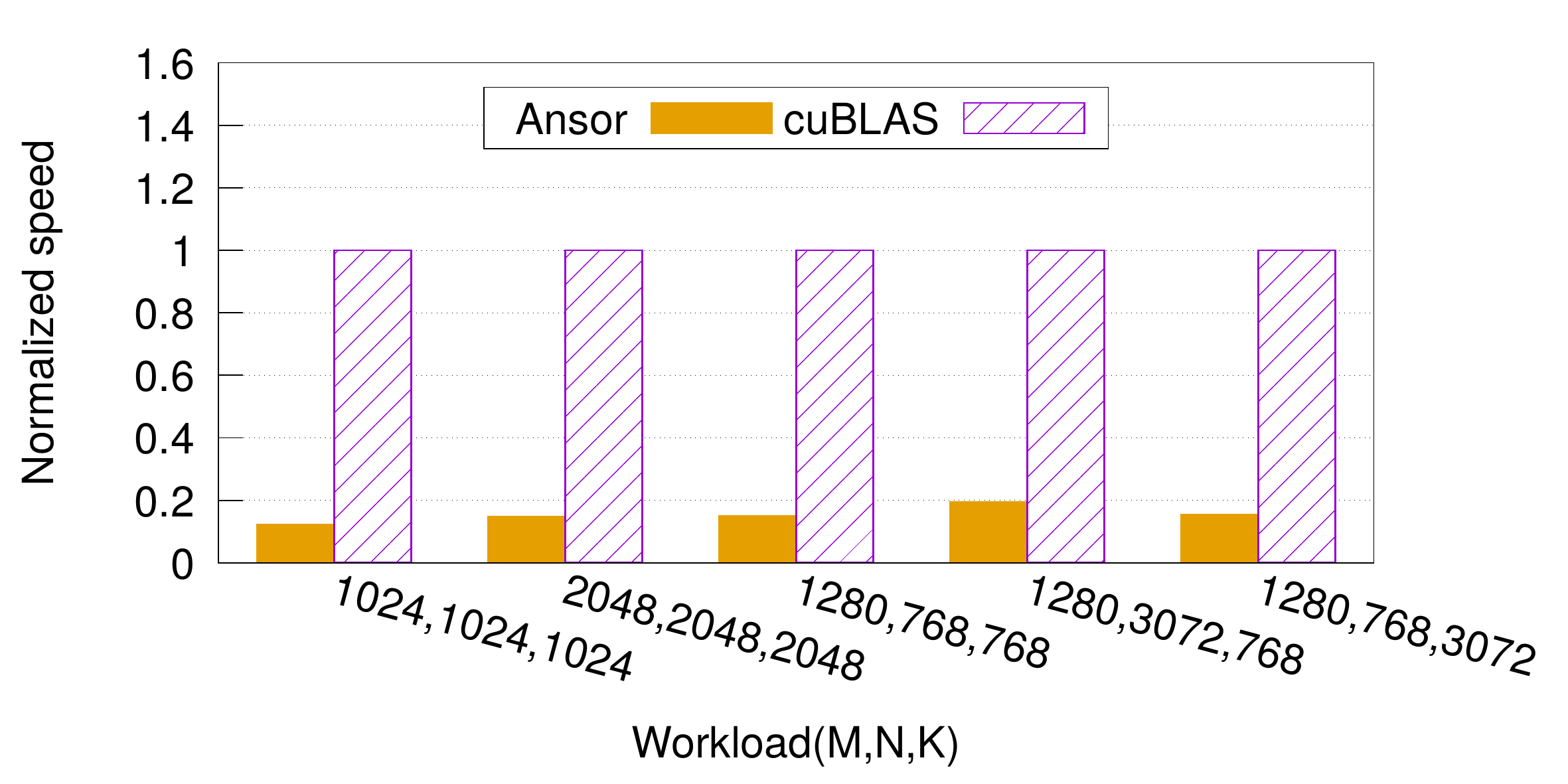}
\caption{The speed of Ansor, implemented in TVM auto-scheduler, under-performs significantly compared to the device speeds achievable in cuBLAS. Workloads: two large square GEMMs and three GEMMs in BERT~\cite{bert}, a widely adopted NLP model, when the batch size is 32 and the sequence length is 40.}
\vspace{-5mm} 
\label{fig:vscublas}
\end{figure}

\textbf{Lack of hardware-native performance.} 
With opaque device models, tensor code generated by existing auto-tuners usually has a performance gap for certain workloads (e.g., non-FP32 GEMM/Conv) as compared to hardware-native performance, as delivered by vendor-tuned libraries, such as cuBLAS and cuDNN. As concrete evidence, Figure~\ref{fig:vscublas} benchmarks the FP16 GEMM speed of tensor programs generated by Ansor~\cite{ansor}, a state-of-the-art auto-tuner, against hardware-native speeds as achieved by cuBLAS. The auto-tuned program achieves less than 20\% of the library performance. The reason is that NVIDIA GPUs have special hardware architecture, tensor cores, to accelerate FP16 computation, but they cannot be efficiently utilized by Ansor that uses an opaque hardware model.

\textbf{Inefficient program search.} 
Opaque device models and inferred hardware execution costs also lead to a less informed tuning process. Existing auto-tuners spend days or weeks when models have many different workloads, e.g., ResNet-152 and Inception-V3~\cite{lorien}. Caching and reusing previous tuning logs~\cite{tophub}) works well for static models, but not those with dynamic data structures~\cite{liang2016semantic} or shapes~\cite{bert}, where the exact workloads are only determined at runtime. In contrast, with hardware-native templated search, \sys reduces the tuning time to tens of minutes for common models.




\subsection{Emerging trend: Templated libraries}

The optimizations in \sys are made possible by an emerging trend: vendor libraries are escaping the earlier generation design with fixed primitives and becoming modularized and composable. Controlled by a set of declarative parameters, templates can be instantiated to suit different hardware and workloads. New primitives can be composed from existing ones, and creating new templates also has a lower barrier. In addition, the templated libraries are efficient design patterns that take into account device details, and extract hardware performance at a level impossible from opaque auto-tuning.


\textit{Example: NVIDIA CUTLASS.} 
Of particular interest to us is CUTLASS, an example templated library from NVIDIA. CUTLASS provides reusable software components in C++ templates for every layer of the CUDA programming model for GEMM. With the right parameters, it achieves high performance for thread-wide, warp-wide, block-wide, and device-wide primitives. Such templates leverage intricate device knowledge, specifically tensor cores as integrated in NVIDIA Volta, Turing, and Ampere GPUs, and optimize for a wide range of mixed-precision computations including B1, INT4, INT8, FP16, BF16, FP32, TF32, FP64, complex, and quaternion. By plugging in the right tile size, data type, and other parameters, users can tap into device-level performance for their workloads.  Beyond this example, Intel and AMD also exhibit similar trends in their design---templated libraries with parameterized control. This design principle, therefore, is generalizable to other platforms. 

\begin{figure}[!t]
    \centering
    \includegraphics[width=7.8cm]{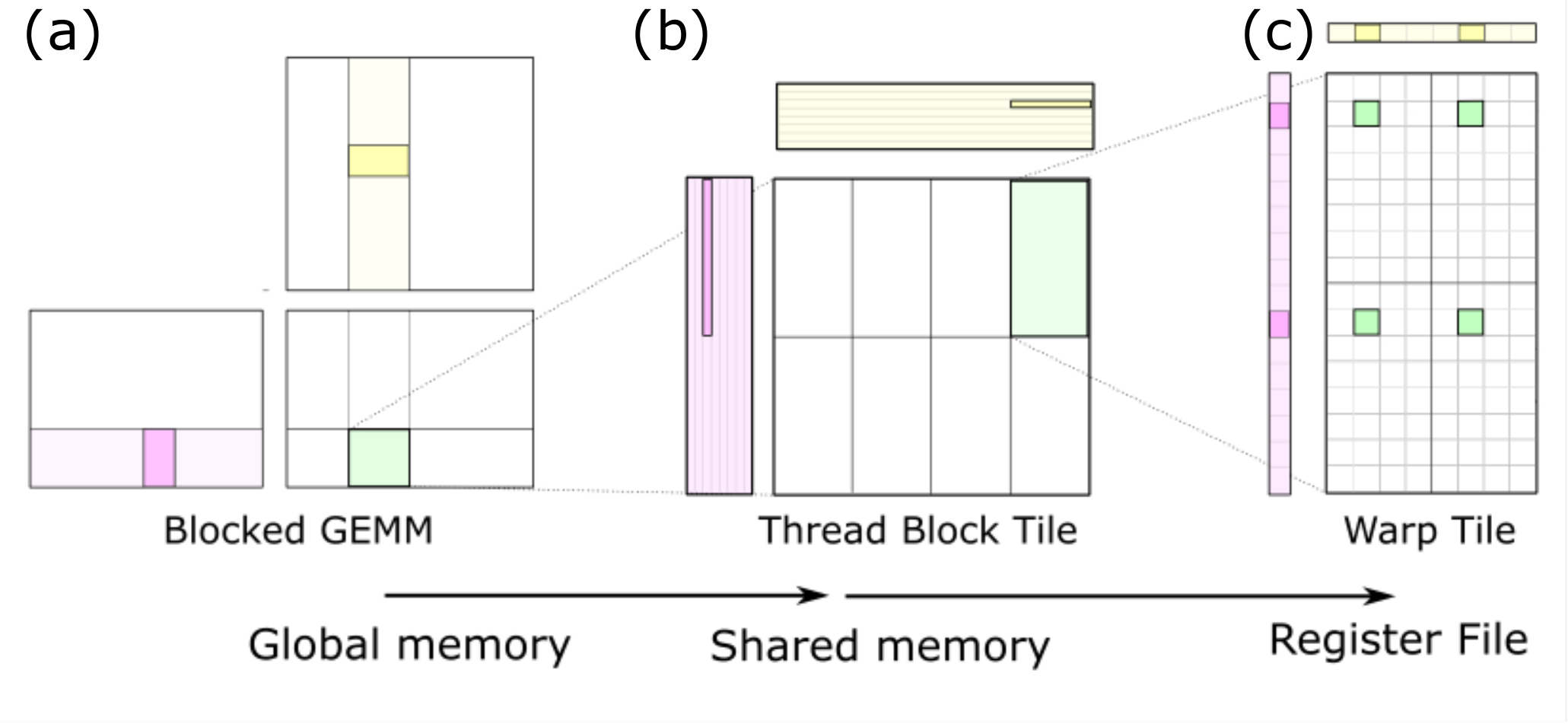}
    \vspace{-3mm}
    \caption{The GEMM hierarchy in CUTLASS and the data movement in threadblock and warp tiles. 
    }
    \label{fig:CUTLASS_hierarchy}
    \vspace{-3mm}
\end{figure}

CUTLASS leverages GPU tiling structures for efficient GEMM implementation by decomposing GEMMs into a hierarchy of threadblocks and warp tiles. It optimizes data movement for locality and carefully controls movement from global to shared memory to the register files. 
Figure~\ref{fig:CUTLASS_hierarchy} illustrates the hierarchy and data movement from slower to faster storage for GEMM operation $C = A\cdot B$. Figure~\ref{fig:CUTLASS_hierarchy}(a) shows the inputs $A$, $B$ and result $C$ in global memory and their threadblock tiles in color (inputs $A$, $B$ in pink and yellow and the result $C$ in green). Threadblock tiles can be divided into warp tiles in shared memory as shown in Figure~\ref{fig:CUTLASS_hierarchy}(b). In this example, a threadblock tile can be split into eight warp tiles which can be further partitioned into thread tiles in the register file as shown in Figure~\ref{fig:CUTLASS_hierarchy}(c). From global memory to shared memory and to register files, the memory size is decreasing but the read/write speed is increasing. Tensor cores on NVIDIA GPUs take thread tiles as input and store the output into register files.

\subsection{\sys: The best of both worlds} 

\sys enables end-to-end optimizations that bridge the gap between auto-tuners and hardware-native performance. 

\textbf{Graph level: Enabling deeper operator fusion.}
Leveraging the templated design, \sys opens up new opportunities for operator optimizations. This is because new optimizations can be introduced to the device libraries via template customization. \sys develops a new operator fusion technique that is called \textit{persistent kernel fusion} for improved performance. Operator fusion computes multiple operators using only one kernel, reducing data shuffling to memory to improve locality~\cite{learningtofuse}, but existing auto-tuner fusions~\cite{tvm, xla, relay, tensorflow, pytorch, taso, learningtofuse} do not interact well with performant, device libraries. For instance, computing Conv2D+BiasAdd+Hardswish in a single kernel improves performance, but the resulting operator may not be supported by fixed-function libraries like cuDNN. Via a templated design, \sys enables new search space that considers deeper fusion, thus opening up graph-level optimizations.

\textbf{Operator level: Automating templated code generation.} 
Templated libraries by themselves, however, are too low-level for common users. Precisely instantiating the parameters to govern tiling sizes and data types creates a high burden. Also, the templated primitives are simply building blocks, and they need to be assembled into complete DNN models for execution. \sys conquers their difficulty of use by combining the library primitives and auto-tuners.
It designs a light-weight performance profiler to search for the best template parameter automatically. By efficiently using the hardware details, the profiler significantly shortens the search time. The search results are later used to instantiates templates and generate the low-level tensor code with hardware-native performance.

\textbf{Model level: System-friendly models.}
The end-to-end optimizations in \sys also shed light on efficient model design. We propose to design models in a system-friendly manner so that they can efficiently use the optimization provided by the underlying systems to achieve better inference performance. In \sys, we have summarized three system-model codesign principles and validated them by augmenting several RepVGG models~\cite{repvgg}.

\section{\sys design}
\label{sec:system} 



\begin{figure}[t!]
\centering\includegraphics[height=5.8cm]{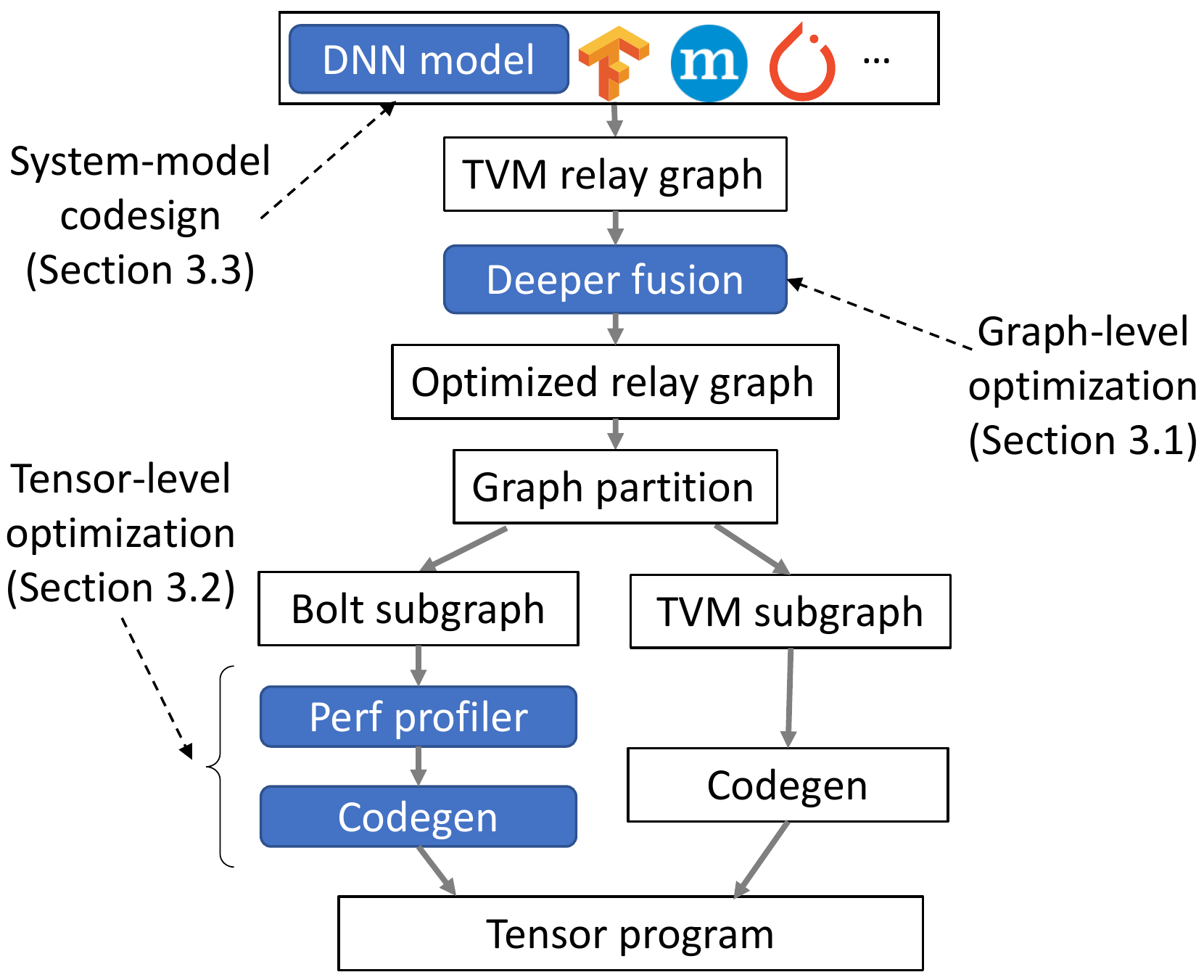}
\caption{The workflow of \sys. Blue boxes are our contributions.} 
\vspace{-5mm} 
\label{fig:overview}
\end{figure}

Figure~\ref{fig:overview} illustrates the workflow of \sys. It follows a BYOC (Bring Your Own Codegen)~\cite{byoc} approach, carving out a suitable subgraph of the tensor program and offloading it to \sys for optimization. Starting from DNN models written in popular frameworks (e.g., TensorFlow, PyTorch, MXNet), \sys reuses the TVM frontend to parse the model into a relay graph. On this graph, it invokes computational graph optimizations (e.g., deeper fusion) and performs graph partition over the optimized graph. \sys next performs hardware-native profiling to search for the best kernels for each operator in the \sys subgraph. Finally, \sys generates high-performance CUDA code which will be compiled together with the code generated by TVM into a single runtime file. In the ensuing discussion, we start with the graph-level, deeper fusion opportunities enabled by \sys, and move down to the automated code generation including the light-weight performance profiler and templated code generator, and finally discuss the system-friendly model design principles distilled from \sys.  


\subsection{Enabling deeper operator fusion} 

\sys enables novel graph-level optimizations by extending hardware-native templates. Specifically, \sys introduces \textit{persistent kernels} which enable novel deeper operator fusion. As shown in Figure~\ref{fig:persistent-kernel-fusion}(a), it works on epilogue fusion as a basis, and further fuses two or more sequential GEMMs/Convs. Fusing multiples GEMMs or Convs into a single operator improves performance in the following ways: (i) eliminating memory traffic for storing and loading inter-layer activations; (ii) eliminating launch latency which is especially beneficial for short kernels with small batch sizes; (iii) enlarging optimization scope for the compiler to explore more instruction scheduling options~\cite{Haicheng:fusion}. Figure~\ref{fig:persistent-kernel-fusion}(b) shows the kernel view of persistent kernel fusion.


\textbf{Prerequisite: Epilogue fusion.} As a prerequisite for persistent kernel fusion, \sys first integrates the epilogue fusion provided in CUTLASS, which fuses a GEMM/Conv kernel with its following epilogues all together into one operator, so that we can further leverage persistent kernels. The epilogue fusion patterns in CUTLASS include: (i) element-wise operators, (ii) data type conversion, (iii) broadcast vector over columns, and (iv) partial reduction over columns. \sys identifies these patterns in the computational graph and generates corresponding algorithmic policy automatically. \sys takes epilogue fusion as a starting point, and 
develops deeper fusions using persistent kernels. 


\subsubsection{Persistent kernel (GEMM/Conv) fusion}
 
\begin{figure}[!t]
    \centering
    \includegraphics[width=0.45\textwidth]{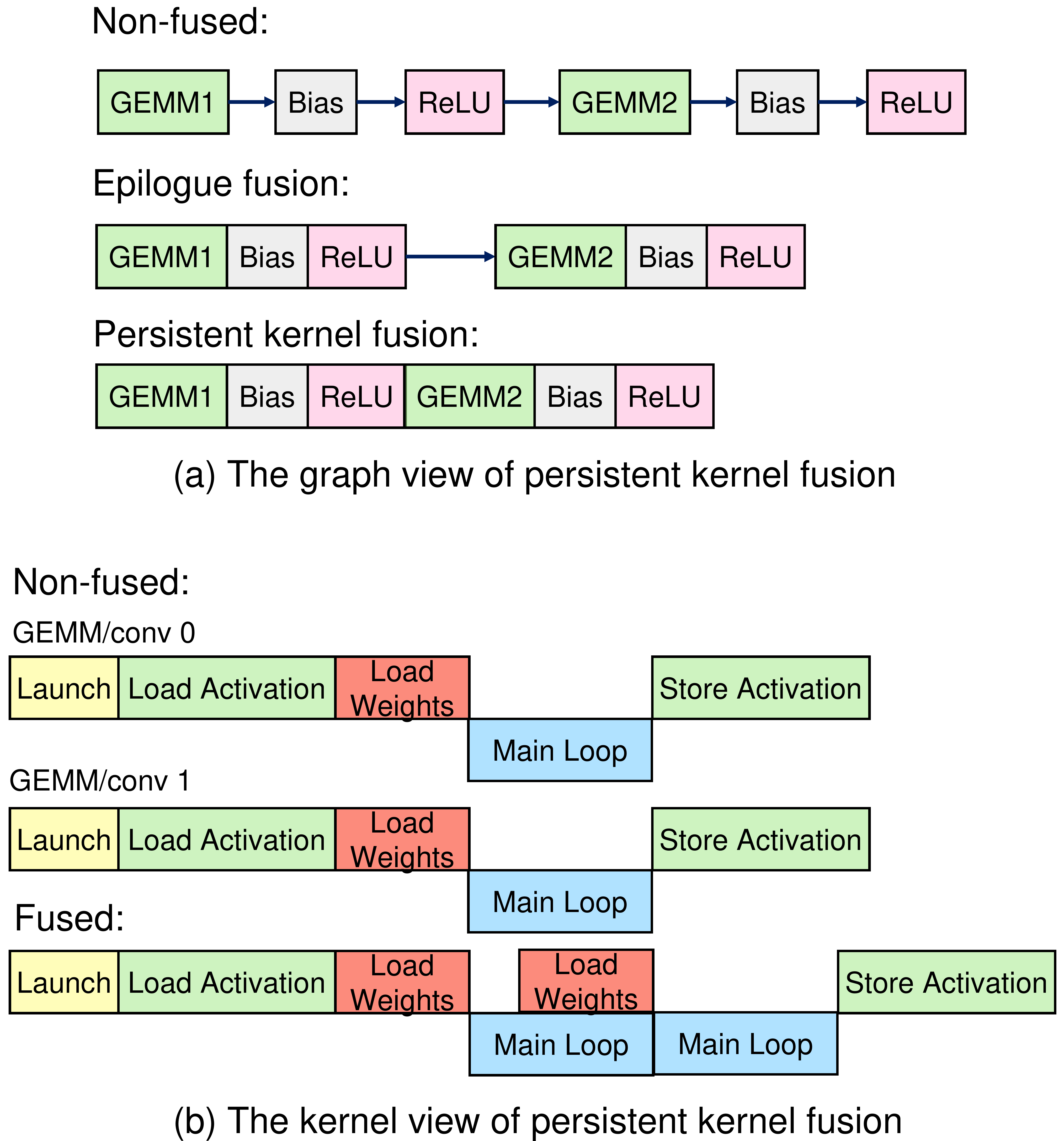}
    \caption{The graph view and kernel view of persistent kernel fusion for back-to-back GEMMs/Convs.}
    \label{fig:persistent-kernel-fusion}
\end{figure}

Persistent kernels allow fusing multiple GEMMs or Convs into one operator to improve performance. As illustrated in Figure.~\ref{fig:persistent-kernel-fusion}(b), when two GEMM/Conv operations are fused together, the main loops of math computation for the two operators run back-to-back in a single fused kernel. The output activation for the first GEMM/Conv stays in faster GPU memory. This eliminates the need for storing GEMM0/Conv0 output activation back to global memory, launching GEMM1/Conv1 kernel, and loading GEMM1/Conv1 input activation from global memory. \sys automatically identifies the opportunity to use persistent kernels and generates CUDA code by creating new templates in CUTLASS. 
We describe the back-to-back GEMM fusion in detail, and convolution fusion works similarly.


A back-to-back GEMM is defined as:
\begin{align}
    D0 = \alpha_0 A0 \cdot W0 + \beta_0 C0,
    \label{EQ:gemm0}
    \\
    D1 = \alpha_1 D0 \cdot W1 + \beta_1 C1,
    \label{EQ:gemm1}
\end{align}
with $A0, W0$ and $W1$ as matrix inputs, $\alpha$s and $\beta$s as scalar inputs, and $C0, C1$ as pre-existing matrices (bias), which will be overwritten by the output.  In order to fuse back-to-back GEMMs, output activation $D0$ of the first GEMM layer must be used as input activation of the second GEMM layer. This requires that the M dimension of the GEMM stays the same for all layers. For back-to-back Convs, this requires that all subsequent Convs (from the 2nd) must use $1\times 1$ filter with no padding and a stride of one.  





\begin{figure}[t]
    \centering
    \includegraphics[width=0.25\textwidth]{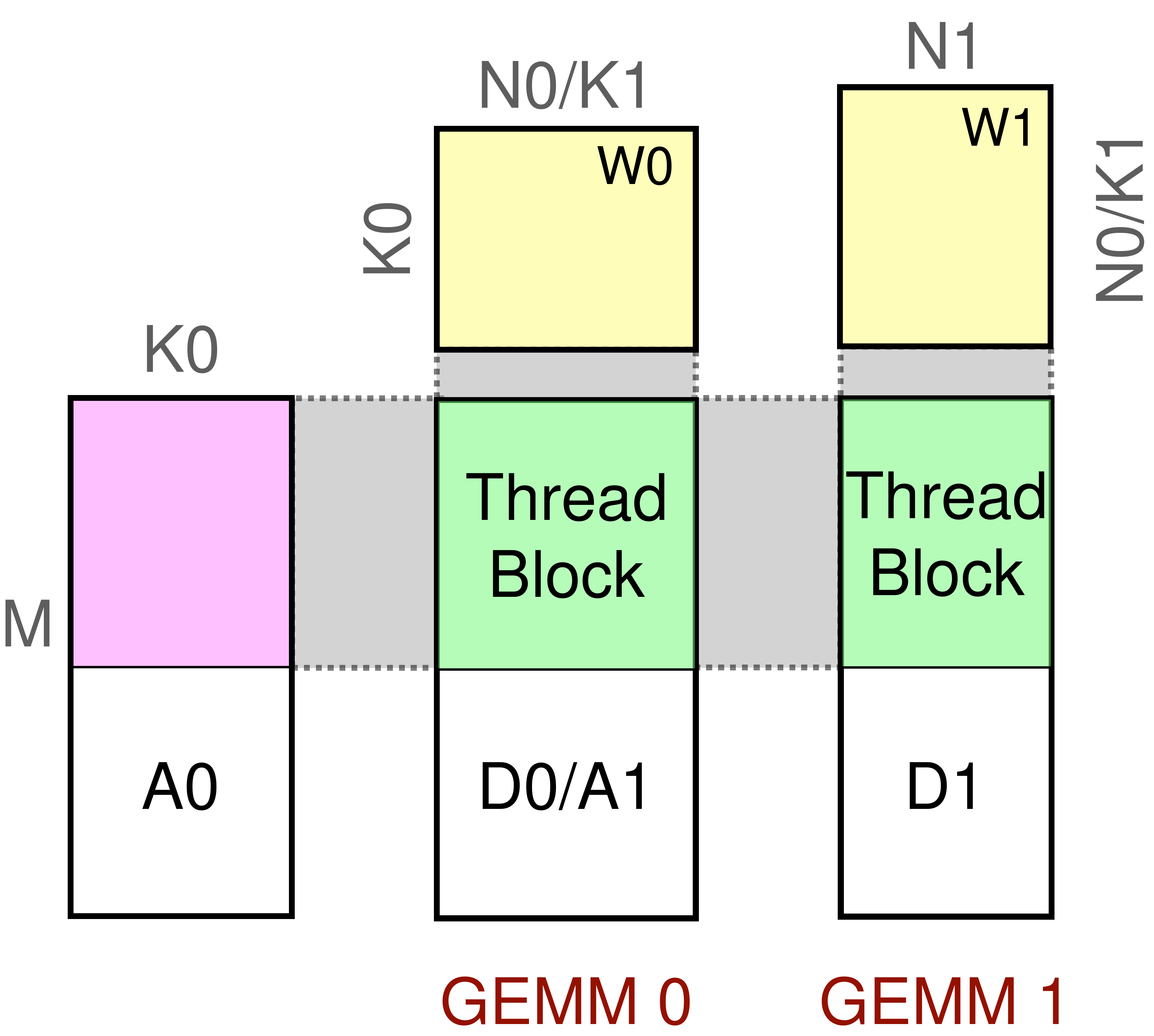}
     \vspace{-2mm}
    \caption{Illustration of threadblock-residence of GEMM fusion. Colored boxes represent one single threadblock. This requires ThreadBlock0\_N = N0, ThreadBlock1\_N = N1.}
    \label{fig:GEMM_TB}
    \vspace{-5mm}
\end{figure}

\textbf{Key property: Threadblock residence.}
The key challenge of persistent kernels is to compute the 2nd GEMM/Conv without loading its input activation from the global memory. This requires each output threadblock of the 1st GEMM/Conv to remain within the same threadblock memory (either in the shared memory or register files) as its respective input threadblock. We call this \textit{threadblock residence}. If it does not hold,  the 2nd GEMM/Conv has to fetch data from the global memory, eliminating the benefits of persistent kernels. For GEMM fusion, threadblock residence requires ThreadBlock\_N = GEMM\_N for each operator. As for Conv fusion, the requirement is ThreadBlock\_N = Conv\_output\_channel. Figure~\ref{fig:GEMM_TB} visualizes this requirement. With threadblock residence, we develop two designs for different scenarios.

\begin{figure}[!t]
    \centering
    \includegraphics[width=0.43\textwidth]{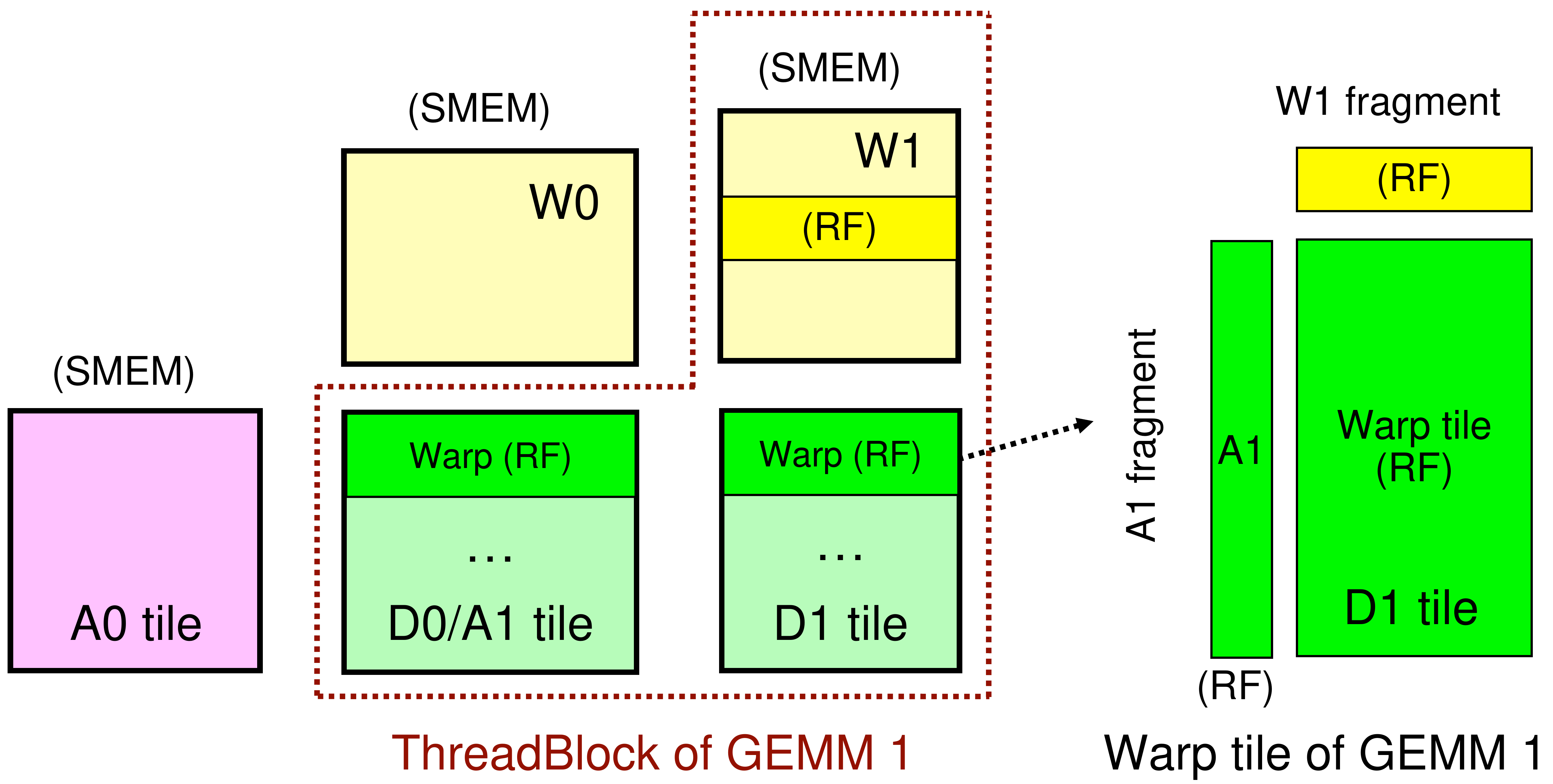}
    \vspace{-2mm}
    \caption{RF-resident fusion in a threadblock of back-to-back GEMMs. The threadblock and warp size requirements are: Warp0\_N=ThreadBlock0\_N=N0, Warp1\_N=ThreadBlock1\_N=N1.}
    \label{fig:RF_resident_fusion}
    \vspace{-4mm}
\end{figure}

\textbf{RF-resident fusion.} 
When the weight matrix $W1$ can be completely streamed into a warp tile in its '$N$' dimension (as indicated in Figure.~\ref{fig:RF_resident_fusion}), threadblock-residence can be satisfied by storing the output activation for each threadblock entirely in the register file (RF). By doing so, the 2nd GEMM/Conv can compute without touching other warps for $W1$. We call this RF-resident fusion which requires that the warp size has to follow Warp\_N = ThreadBlock\_N = GEMM\_N for each layer. In RF-resident fusion, each warp will own a chunk of the accumulator data in the RF (referred to as accumulator fragment) produced by the current layer. This will be used entirely as the input for the next layer computed by the same warp. We develop a CUTLASS warp fragment iterator to extract the data from the accumulator fragment and feed it into warp-level MMA operations. RF-resident fusion incorporates back-to-back MMA pipelines by extending the threadblock-level GEMM design in CUTLASS. Our design has no interference across GEMM operations. The only extra operation for the 2nd GEMM is to get warp fragments from the previous accumulator and perform epilogue computation all in the RF.

\begin{figure}[!t]
    \centering
    \includegraphics[width=0.43\textwidth]{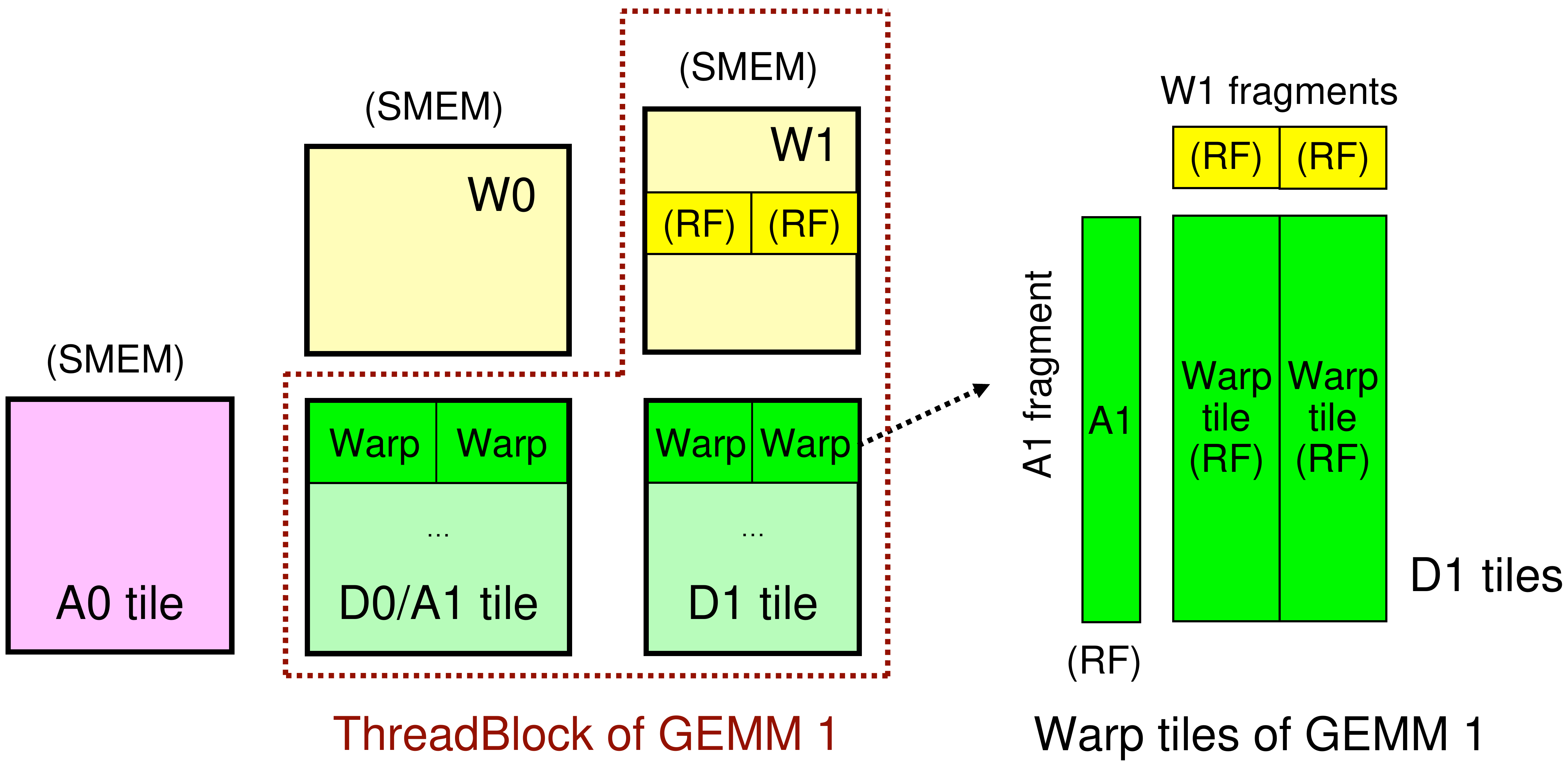}
    \caption{Shared memory-resident fusion in a threadblock of back-to-back GEMMs. The threadblock size requirements are: ThreadBlock0\_N = N0 $\neq$ Warp0\_N, ThreadBlock1\_N = N1 $\neq$ Warp1\_N.}
    \label{fig:shmem_resident_fusion}
    \vspace{-6mm}
\end{figure}

\textbf{Shared memory-resident fusion.} 
RF-resident GEMM fusion creates higher RF pressure especially when GEMM\_N is large, which will potentially harm the kernel performance and limit the applicable scenarios. To solve the problem, we propose shared memory-resident fusion to relax the warp size restriction. In this design, when the 2nd GEMM/Conv requires data sharing between warps, the data can be staged into shared memory instead of RF. Figure~\ref{fig:shmem_resident_fusion} shows an example where the computation for $D1$ has to stream $W1$ fragments from multiple warp tiles in the '$N$' dimension. Thus, the accumulator data produced in GEMM0 must be transferred from RF to shared memory in order to be loaded by GEMM1. The data chunk owned by each warp will be shared in $M$ dimension for the next layer. By doing so, the warp size restriction of Warp\_N in RF-resident fusion can be relaxed. 
To enable shared memory-resident fusion, we introduce a smem fragment iterator as the mechanism to store the accumulator tile into shared memory, and then fetch fragment from shared memory for the 2nd GEMM. In order to achieve higher performance, we carefully design the shared memory layout to avoid any shared memory bank conflict when storing the accumulators of the 1st kernel and loading it for the 2nd one.

\textbf{Summary.} RF-resident and shared memory-resident fusion enables deeper fusion of sequential GEMMs/Convs. Based on the back-to-back fusion, \sys can support fusing multiple  GEMMs/Conv by extending the persistent kernel templates and duplicating the GEMM pipelines.


\subsection{Automating templated code generation}
\label{sec:tvm} 

\subsubsection{Challenges in code generation} 

Templated libraries pose new challenges for end-to-end tensor program optimization. Foremost, 
these templates usually do not provide complete functionality for end-to-end models, but only support a subset of operators. 
One na\"ive solution is to develop a full compiler stack from scratch for each hardware, but this does not scale. 
\sys addresses this challenge by employing a BYOC (Bring Your Own Codegen)~\cite{byoc} approach. It enables us to reuse the existing compiler stacks (e.g., TVM) as much as possible and focus only on the optimization and code generation using templated device libraries.

A na\"{i}ve application of BYOC does not solve the entire problem. First, templated device libraries by themselves are not directly runnable. They require users to instantiate the template with well-tuned parameters to achieve good performance, but BYOC does not support such performance profiling. \sys addresses the problem by proposing a light-weight hardware-native performance profiler that can search for the best parameters for an operator with a specific workload within minutes. 
In addition, conventional BYOC regards device libraries as agnostic external functions and generates hardware code with hooks to invoke them at runtime. 
This design makes it difficult to customize the hardware library and support new optimizations, such as layout transformation and kernel padding. \sys resolves the problem by viewing the library as a whitebox and generating code in its convention directly. In the following, we will describe our detailed design. 


\subsubsection{Light-weight performance profiler} 

\sys designs a light-weight performance profiler to search for the best template parameters. Traditional auto-tuners, assuming no hardware information, infer a cost model via generating sample implementations and measuring their speeds, which requires auto-tuners to explore a large search space and leads to long tuning time. \sys greatly reduces the search time by separating the time-consuming sample program generation from performance measurement, and by effectively using the hardware details to accelerate the former. Specifically, the performance-related parameters in CUTLASS templates include threadblock, warp, and instruction shapes, swizzling functor, stages, etc. \sys determines their possible values according to the GPU architecture as well as tuning guidelines that are specific to each hardware, thanks to the whitebox approach. For example, within the capacity of register files, \sys prefers large warp tile sizes to achieve a higher compute-memory ratio; four or eight warps per threadblock tends to have better performance when running on modern NVIDIA GPUs; small problem sizes need small threadblock sizes to launch enough thredablocks to keep more SMs busy. For each GPU architecture, \sys produces tens of best parameter combinations and generates the corresponding sample programs by initiating the template. Note that these sample programs are reusable across models and workloads by given different inputs. Therefore, at runtime, \sys can profile the performance by calling the pre-generated sample programs with concrete inputs.

\subsubsection{Templated code generation}
\label{sec:codegen}
Traditional BYOC systems~\cite{byoc} cannot target code generation in templated format; they treat such libraries as external functions at runtime. In contrast, \sys produces low-level tensor implementations in the CUTLASS  convention by instantiating the templates with the best parameters identified by the profiler. Our approach has two advantages. First, the generated code delivers superior performance, e.g., can reach 300 TFLOPS throughput for FP16 GEMM on Ampere A100 which is more than 95\% of the hardware theoretic limit. Second, it provides full flexibility to add novel optimizations in the generated code. In \sys, we develop the following two optimizations.

\textbf{Layout transformation.} CUTLASS supports only NHWC layout for Convs because it is faster than NCHW layout~\cite{layout}. But not all models are written in the desired layout---e.g., all Pytorch models use NCHW. To enable more optimization opportunities, \sys provides automated layout transformation. Note that this is different from the layout transformation provided in TVM that achieves the function by modifying the relay graph. Rather, \sys implements the transformation in the generated CUDA code of the model's first and last layer directly to save extra kernel launch overhead. The transformation requires a new tensor to hold the data with the new layout. Allocating and imitating the new tensor within the kernel will create significant memory overhead. Instead, we pre-allocate the memory by adding a new variable in the model's parameters that can be used by the kernel directly.

\textbf{Kernel padding.}
Although CUTLASS supports alignments 8, 4, 2, 1 to cover all different workloads, the performance varies significantly across different alignments. The largest vectorized load and store supported by NVIDIA GPUs is 128 bits, so the most efficient way to use it for FP16 data type is alignment 8 (128/16). Using alignment 8 in this case can reduce the load and store instruction counts, as well as the number of predicates needed by every load and store instruction. Tensor shapes with a dimension that cannot be divided by 8 will have to use smaller alignments. For instance, the first layer of convolutional neural networks usually has three input channels, which has to use alignment 1. This will suffer from non-coalesced memory access and shared memory bank conflicts. Therefore, \sys automatically pads unaligned tensors to use alignment 8. It allows us to not only fully utilize tensor core acceleration, but also to reduce memory loading time. Similar to layout transformation, we also pre-allocate the aligned tensor memory in models' parameters.

\subsection{Designing system-friendly models}

The graph-level optimization (e.g., persistent kernel fusion) and operator-level optimization (e.g., automated padding) in \sys also shed light on model-level optimization opportunities. Models that are designed in a way that effectively makes use of the system strengths can lead to more efficient inference. We call this \textit{system-model codesign}, which can help build system-friendly models running more efficiently. 
\sys identifies the following principles for this codesign.  

\textbf{Exploring different activation functions with epilogue fusion.}
The selection of activation functions has a notable influence on the accuracy of DNN models~\cite{swish}. Over the years, a line of activation functions have been designed, such as ReLU~\cite{relu} and its variants, GELU~\cite{gelu}, Softplus~\cite{softplus}, and Hardswish~\cite{hardswish}. 
In \sys, the epilogue fusion will fuse activations with the leading GEMM/Conv to reduce the overhead 
of activations. Therefore, model designs could explore different activation functions in their models 
and identify the most effective one. 


\textbf{Deepening models with 1$\times$1 Convs.} 
Deepening neural networks to achieve higher accuracy is a commonly-used model design technique. For instance, ResNet~\cite{resnet} has different depths from 18 layers to 151 layers with increasing accuracy. However, the inference speed will drop quickly as the depth increases. Deepening models with 1$\times$1 Convs, on the other hand, only incurs low computation overhead in \sys. This is because of the persistent kernel fusion optimization. Therefore, although deepening models with 1$\times$1 Convs does not increase accuracy to the same extent as larger kernels, one can still add 1$\times$1 Convs to improve the accuracy with reduced speed loss.


\textbf{Aligning tensor shapes to use GPUs more efficiently.}
As we discussed in Section~\ref{sec:codegen}, tensor shapes have significant impacts on the efficiency of models running on GPUs. Although \sys will automatically perform padding over unaligned tensors, the padding itself will incur extra overhead, as shown in Table~\ref{tab:pad}. As a result, one could design models with aligned tensor shapes to achieve higher efficiency, avoiding additional padding that will be needed. 

\section{Evaluation}
\label{eva}

Our evaluation of \sys focuses on the following aspects. First, we perform microbenchmarks to evaluate the performance of \sys in terms of GEMM/Conv2D computation, epilogue fusion, persistent kernel fusion, and kernel padding. Second, we evaluate the end-to-end performance of \sys on widely-used convolutional neural networks. Finally, we apply our system-model codesign principles to the RepVGG models~\cite{repvgg} as a case study.

\textbf{Setup.} Our experiments are performed on a single NVIDIA Tesla T4 GPU. We use Ansor~\cite{ansor}, the state-of-the-art auto-tuner in TVM as our baseline. All inference computations in the evaluation use the FP16 data type.

\subsection{Microbenchmarks}

\subsubsection{GEMM/Conv2D performance}

\begin{figure*}[t]
\centering
\subcaptionbox{GEMMs performance.\label{fig:gemm}}%
      [.4\linewidth]
      {\includegraphics[height=4cm]{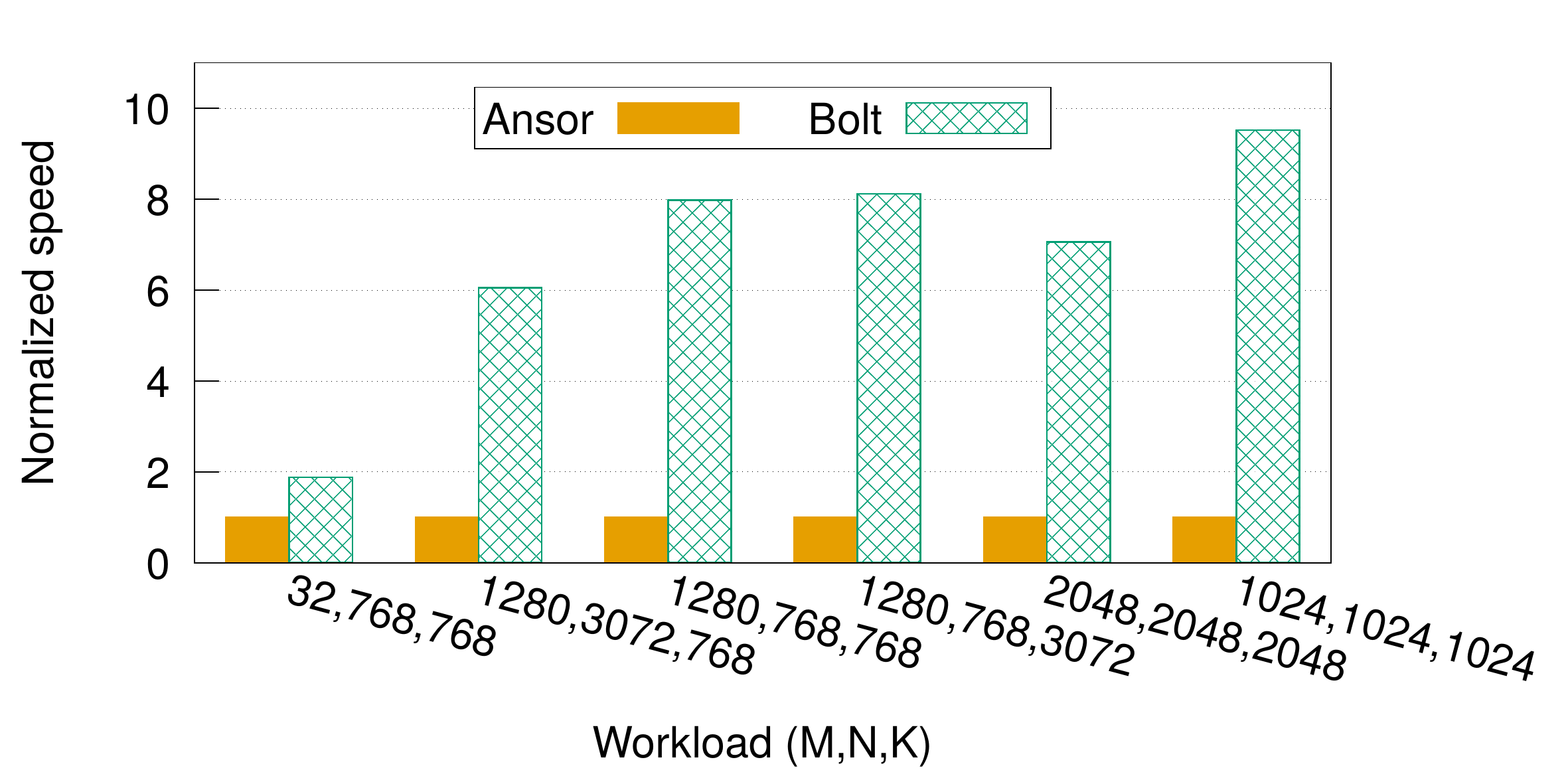}}
\hspace{3em}%
\subcaptionbox{Conv2D performance. \label{fig:conv}}%
      [.4\linewidth]
      {\includegraphics[height=4cm]{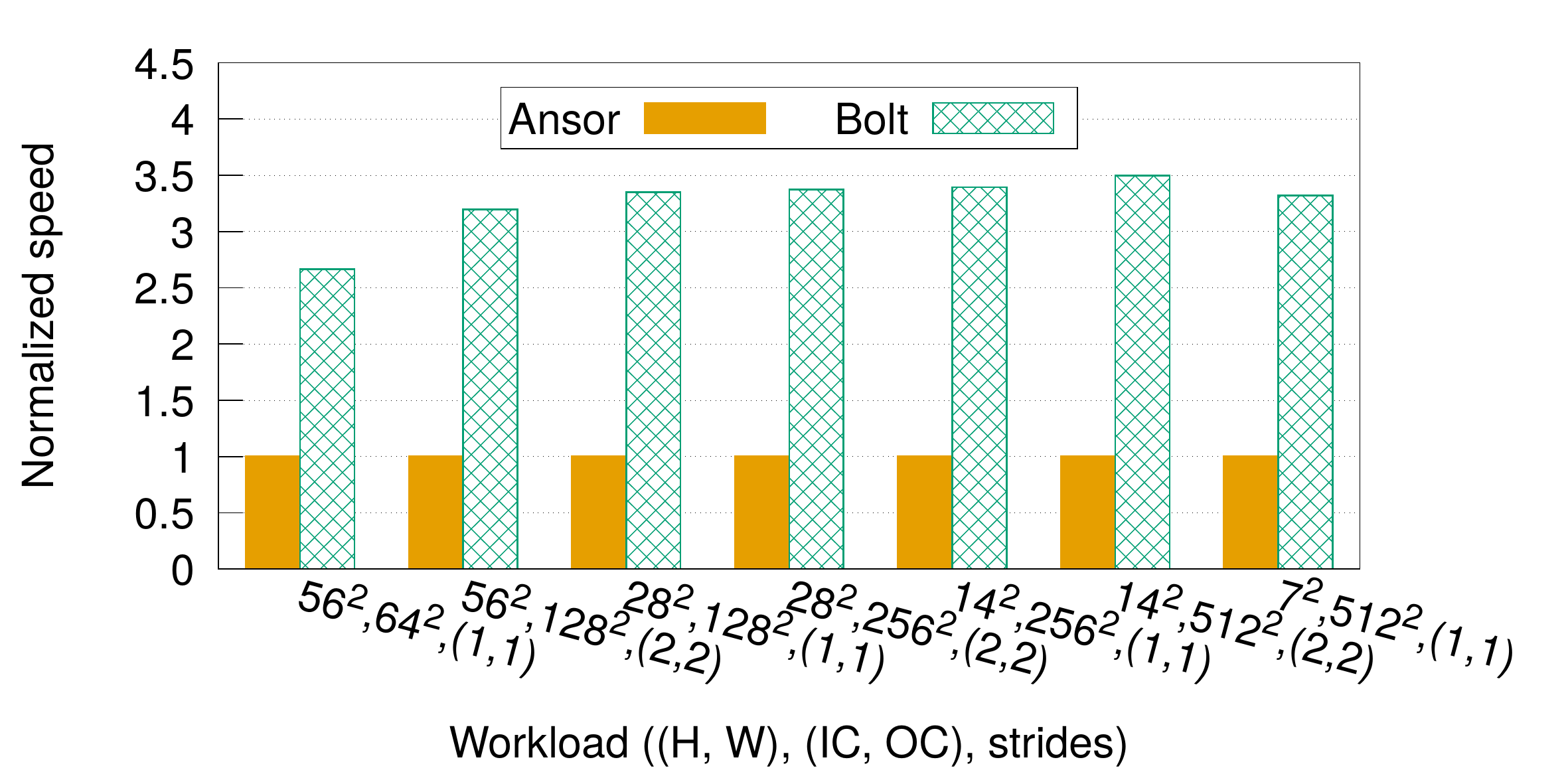}}
      \vspace{-2mm}
\caption{The performance of \sys on GEMMs and Conv2Ds. Figure~\ref{fig:gemm} shows the speed of GEMMs in BERT with batch size=32 and sequence length=40 and two square GEMMs. Figure~\ref{fig:conv} shows the speed of $3\times3$ Conv2Ds in ResNet-50. The batch size=32 and all Conv2Ds use (1, 1) zero padding.}
\vspace{-4.5mm}
\label{fig:gemm_conv}
\end{figure*}

We first evaluate the performance of \sys-generated GEMM and Conv2D implementations. For GEMMs, we evaluate a) typical GEMMs in BERT~\cite{bert} where the batch size is 32 and sequence length is 40 and b) two square GEMMs. For the Ansor baseline, we tune each workload for 2000 trials for performance optimization, following the TVM official example. We run each workload 1000 times and compute the average speed; results are shown in Figure~\ref{fig:gemm}. Ansor adopts a strategy that aggressively consumes all register files to achieve higher performance. However, this greedy approach is only effective for less compute-intensive workloads. Therefore, \sys is 6.1-9.5x faster than Ansor on compute-intensive workloads and achieves 1.9x speedup on the one that is less compute-intensive. 
Similarly, we measure the speed of Conv2Ds in Figure~\ref{fig:conv}. The workloads are extracted from ResNet-50 using 32 as the batch size. All Conv2Ds in the table are using (3, 3) kernels and (1, 1) zero padding. For all cases, \sys is 2.7-3.5x faster than Ansor.
Overall, \sys achieves significantly higher performance as the tuning strategy based on hardware-native templates
extracts native performance.

\subsubsection{Epilogue fusion performance}

\begin{figure*}[t]
\centering
\subcaptionbox{GEMM epilogue fusion.\label{fig:epilogue_gemm}}%
      [.4\linewidth]
      {\includegraphics[height=4cm]{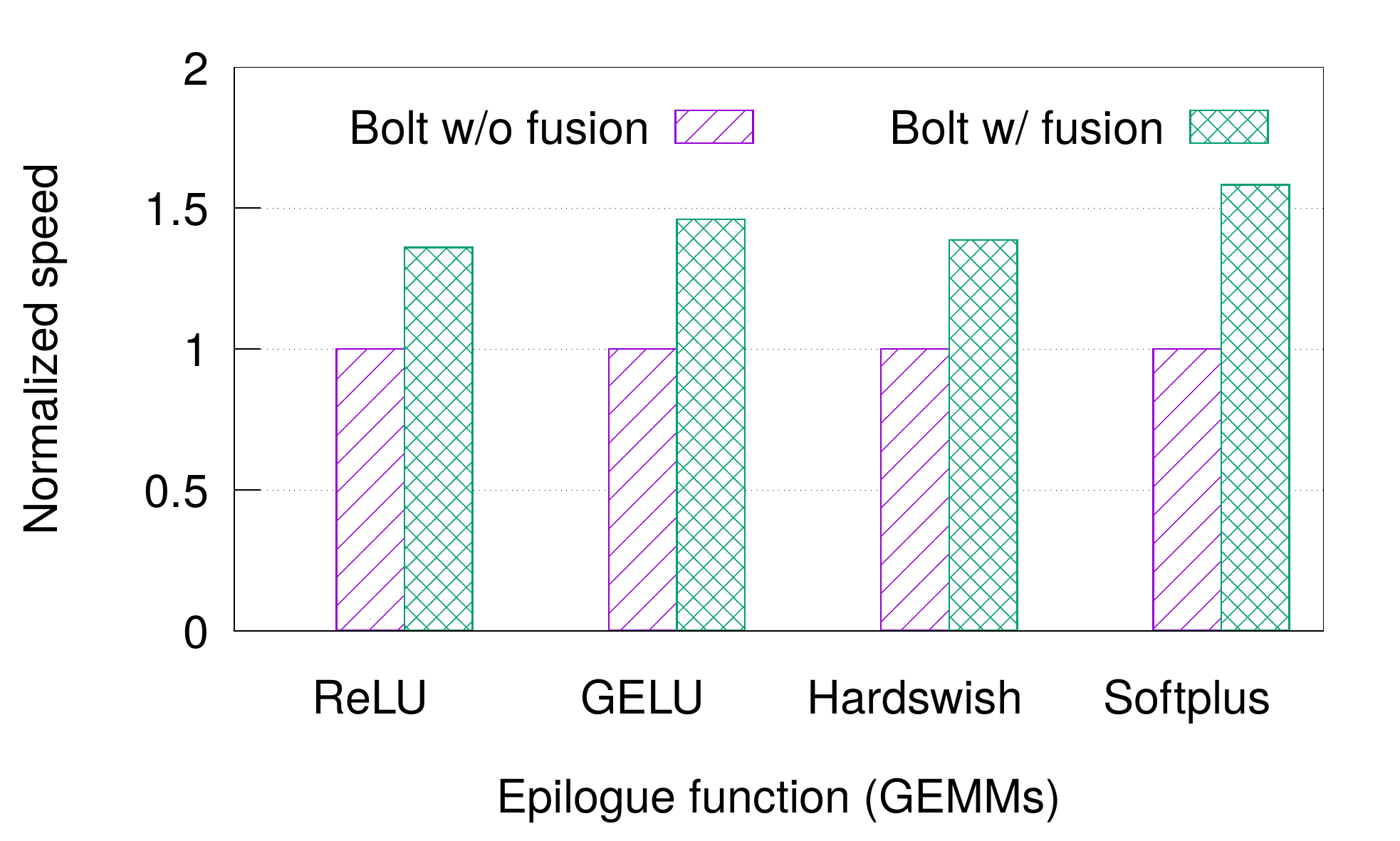}}
\hspace{3em}%
\subcaptionbox{Conv2D epilogue fusion.\label{fig:epilogue_conv}}%
      [.4\linewidth]
      {\includegraphics[height=4cm]{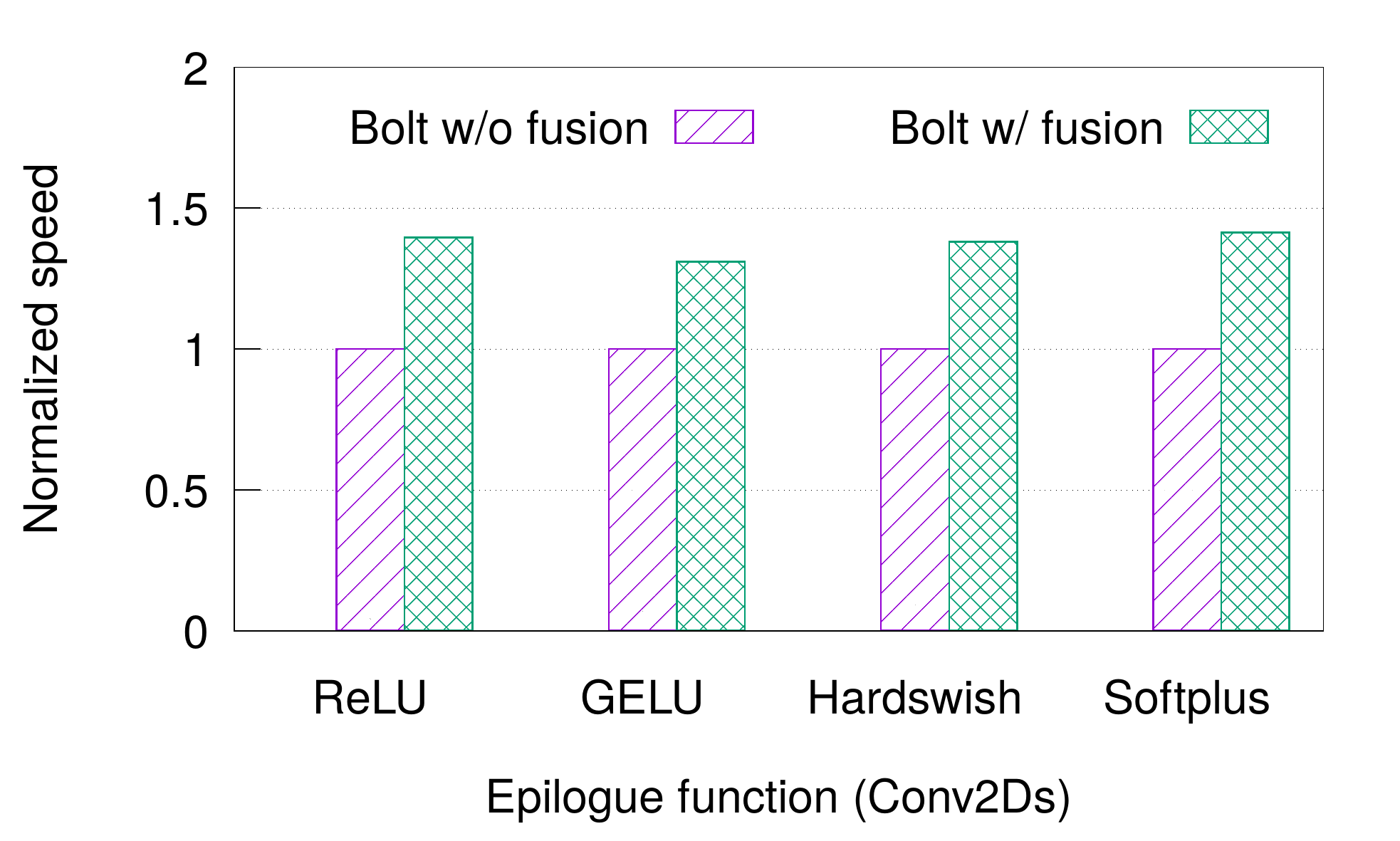}}'
      \vspace{-2mm}
\caption{The performance of epilogue fusion on pattern GEMM/Conv2D+BiasAdd+Activation. The workload of the GEMM is M=1280, N=3072, and N=768. The workload of the Conv2d is H=W=56, IC=OC=64, kernel=(3, 3), stride=(1,1), and padding=(1,1).}
\vspace{-3.5mm}
\label{fig:epilogue}
\end{figure*}

We then evaluate the effectiveness of epilogue fusion on element-wise operators. We choose one workload for GEMM and Conv2D from Figure~\ref{fig:gemm_conv} respectively and measure the performance of the pattern GEMM/Conv2D+BiasAdd+Activation. We experiment on four different activation functions: ReLU, GELU, Hardswish, and Softplus, and the results are shown in Figure~\ref{fig:epilogue}. Our baseline here is \sys without epilogue fusion, in which \sys only computes the GEMM/Conv2D and TVM will fuse the BiasAdd and activation and compute them as one operator. 
As we can see, epilogue fusion improves the computation speeds for both GEMMs and Conv2Ds. 
The average speedup for GEMMs and Conv2D is 1.45x and 1.38x respectively. We have observed similar performance gains on other workloads (not shown).

\subsubsection{Persistent kernel fusion performance}

We next evaluate the performance of persistent kernel fusion. First, we use \sys to fuse two GEMMs with the pattern GEMM1+ReLU+GEMM2+ReLU into one GEMM operator using RF-based or shared-memory based persistent kernels, depending on their performance. The baseline is \sys with only epilogue fusion that computes the two GEMMs sequentially. Results are presented in Table~\ref{tab:persistent-gemm}. Workloads are extracted from real recommendation models, e.g., DCNv2~\cite{dcn}, DLRM~\cite{DLRM}.
As we can see, the persistent kernel fusion accelerates the computation by 1.2x-1.5x. 
For Convs, we extract the $3\times3$ Conv2Ds from the first a few layers in the RepVGG models and create a $1\times1$ Conv2D for each of them.  
As shown in Table~\ref{tab:persistent-conv}, our persistent kernel fusion can improve the computation speed by 1.1x-2.0x. Note that our persistent kernels can fuse more than two GEMMs/Convs, which can further improve the performance by saving intermediate memory access and kernel launch. 

\begin{table}[!t]
\centering
\caption{The performance of fusing two back-to-back GEMMs using persistent kernels. Each GEMMs is followed by a ReLU epilogue and all of them will be fused into one kernel.}
\vspace{1mm}
\scalebox{0.83}{
\begin{tabular}{p{10mm}p{3mm}p{5mm}||p{10mm}p{3mm}p{5mm}||cc}
\Xhline{3\arrayrulewidth}
\multicolumn{3}{c||}{1st GEMM} & \multicolumn{3}{c||}{2nd GEMM}  & \multicolumn{2}{c}{Normalized speed}  \\ \Xhline{2\arrayrulewidth}
M & N & K & M & N & K & w/o fuse. & w/ fuse. \\ \Xhline{2\arrayrulewidth}
2464    & 1     & 4     & 2464      & 4     & 1     & 1.00  & 1.24   \\ 
16384   & 64    & 256   & 16384   & 16    & 64    & 1.00  & 1.34   \\ 
32768   & 128   & 576   & 32768   & 64   & 128    & 1.00  & 1.28    \\
128320    & 32     & 96     & 128320      & 96     & 32     & 1.00  & 1.46   \\ \Xhline{3\arrayrulewidth}
\end{tabular}
}
\label{tab:persistent-gemm}
\vspace{-5mm}
\end{table}

\begin{table}[!t]
\centering
\caption{The performance of fusing two back-to-back Conv2Ds using persistent kernels. Each Conv2D is followed by a BiasAdd and a ReLU epilogue. The $3\times3$ Conv2D uses (1, 1) padding and the $1\times1$ Conv2D uses (1, 1) strides and does not have padding.}
\vspace{1mm}
\scalebox{0.83}{
\begin{tabular}{p{8mm}p{10mm}p{8mm}||p{8mm}p{10mm}||cc}
\Xhline{3\arrayrulewidth}
\multicolumn{3}{c||}{3$\times$3 Conv2D} & \multicolumn{2}{c||}{1$\times$1 Conv2D} & \multicolumn{2}{c}{Normalized speed} \\ \Xhline{2\arrayrulewidth}
H, W & IC, OC & strides & H, W & IC, OC &  w/o fuse. & w/ fuse.  \\ \Xhline{2\arrayrulewidth}
$224^{2}$ & 3, 48 & (2, 2) & $112^{2}$ & 48, 48  & 1.00   &1.10    \\ 
$112^{2}$ & 48, 48 & (2, 2) & $56^{2}$ & 48, 48   & 1.00  &1.41    \\ 
$56^{2}$ & 48, 48 & (1, 1) & $56^{2}$ & 48, 48  &    1.00  & 1.87   \\ 
$224^{2}$ & 3, 64 & (2, 2) & $112^{2}$ & 64, 64 & 1.00  & 1.24  \\ 
$112^{2}$ & 64, 64 & (2, 2) & $56^{2}$ & 64, 64 &   1.00  & 1.12   \\ 
$56^{2}$ & 64, 64 & (1, 1) & $56^{2}$ & 64, 64  &    1.00  & 2.02 \\ \Xhline{3\arrayrulewidth}
\end{tabular}
}
\label{tab:persistent-conv}
\vspace{-5mm}
\end{table}

\subsubsection{Padding performance and overhead}

We now evaluate the performance benefits of automated padding and its extra overhead in \sys. In Table~\ref{tab:pad}, we choose a few Conv2D workloads in our production of which the input channels are not divisible by eight. Without padding, these workloads can only compute with alignment 2. \sys will automatically pad them to the closest 8-divisible sizes, thus leveraging alignment 8. We measure the normalized computation speed of \sys with and without padding. As we can see, after padding, the speed can be improved by 1.8x on average. However, the padding itself will incur extra overhead: in this benchmark, the average time spent on padding over the total computation time (padding+Conv2D computation) is 16\%. This is further evidence of our 3rd system-friendly model design principle---models should be designed with aligned tensor shapes.

\begin{table}[!t]
\centering
\caption{The performance and overhead of \sys's automated padding. Unpadded Conv2Ds are computed with alignment=2; after being padded, alignement=8 can be used. The cost of padding is the time spent on the padding over the total computation time (padding+Conv2D).}
\vspace{1mm}
\scalebox{0.83}{
\begin{tabular}{p{3mm}cp{11mm}cc||p{6mm}p{5mm}||p{5mm}}
\Xhline{3\arrayrulewidth}
\multirow{2}{*}{N} & \multirow{2}{*}{H, W} & \multirow{2}{*}{IC, OC} & \multirow{2}{*}{kernel} & \multirow{2}{*}{padding} & \multicolumn{2}{c||}{Norm. speed} & \multirow{2}{*}{Cost}\\ \cline{6-7} 
 &  &  &  &  & unpad & pad & \\ \Xhline{2\arrayrulewidth}
32 & 20, 26 & 46, 32 & (3, 3) & (1, 1) & 1.00 & 1.62  & 18\% \\ 
32 & 20, 26 & 46, 32 & (5, 5) & (2, 2) & 1.00 & 1.95  & 9\%\\ 
128 & 14, 19 & 46, 32 & (5, 7) & (0, 0) & 1.00 & 1.77 & 15\%\\
288 & 11, 15 & 46, 32 & (5, 7) & (0, 0) & 1.00 & 1.71 & 18\%\\ 
32 & 20, 26 & 174, 64 & (3, 3) & (1, 1) & 1.00 & 1.60 & 24\%\\ 
32 & 20, 26 & 174, 64 & (5, 5) & (2, 2) & 1.00 & 1.99 & 12\%\\ \Xhline{3\arrayrulewidth}
\end{tabular}
}
\label{tab:pad}
\vspace{-5mm}
\end{table}

\subsection{End-to-end optimization}

\begin{figure*}[t]
\centering
\subcaptionbox{Inference speed.\label{fig:e2e_speed}}%
      [.4\linewidth]
      {\includegraphics[height=4cm]{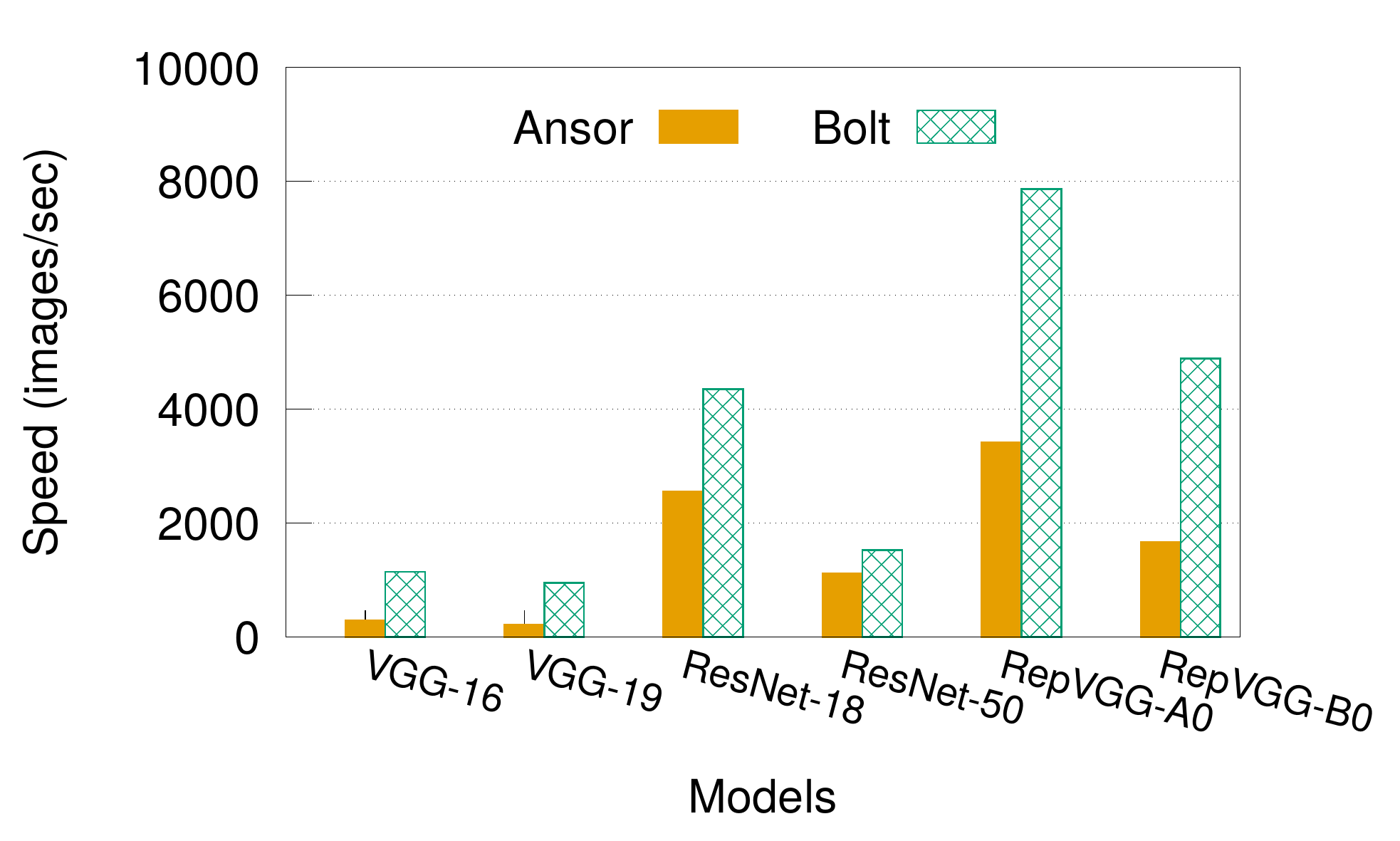}}
\hspace{3em}%
\subcaptionbox{Tunning time.\label{fig:e2e_tuning}}%
      [.4\linewidth]
      {\includegraphics[height=4cm]{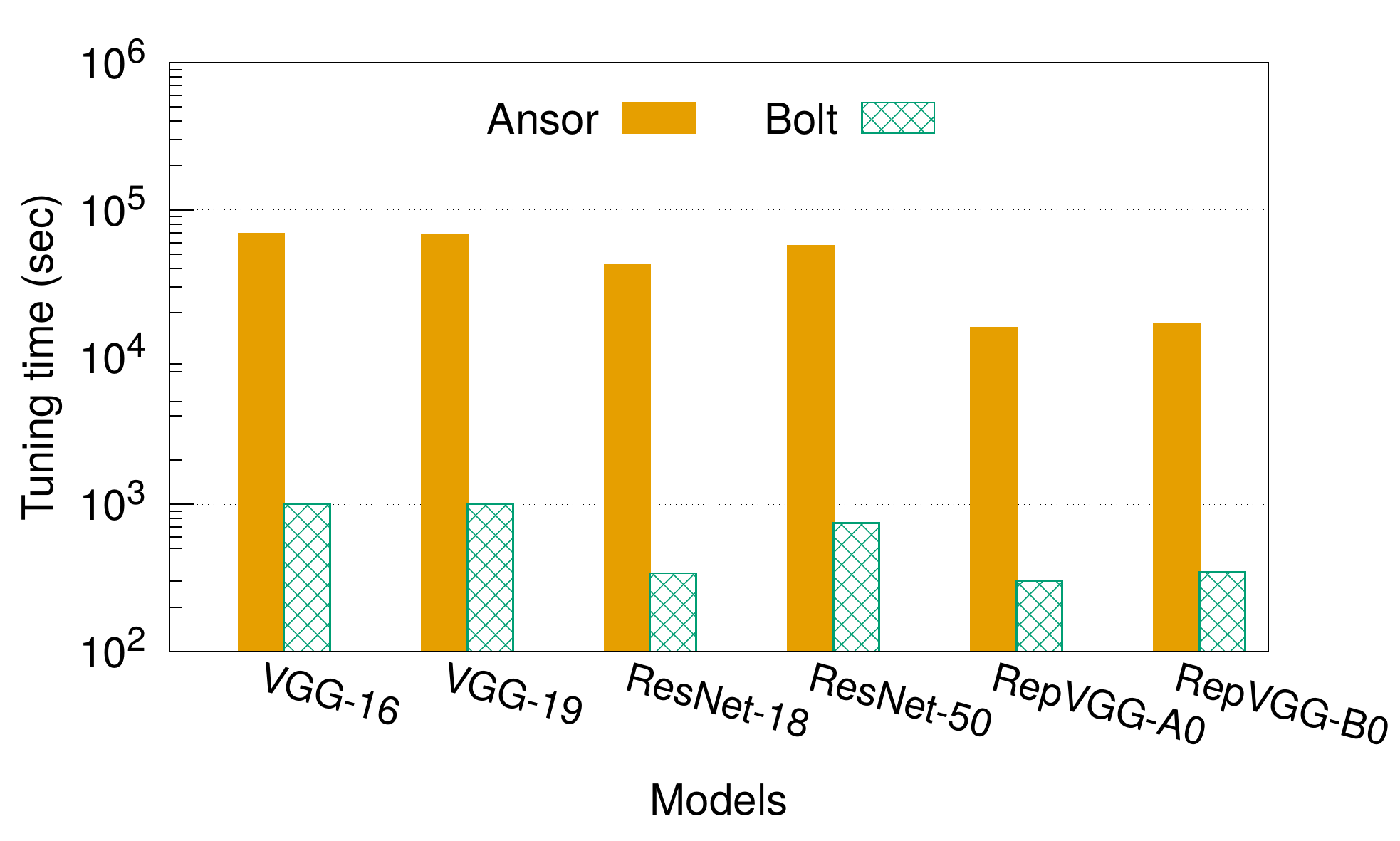}}
\caption{The normalized inference speed and tuning time for widely used convolutional neural networks.}
\vspace{-3.5mm}
\label{fig:e2e}
\end{figure*}

We evaluate the performance of \sys on end-to-end model optimization by experimenting on six widely-used convolutional neural networks. Our baseline is Ansor which performs auto-tuning to optimize performance. We configure Ansor following the official example and set the tuning trials to the recommended 900 $\times$ the number of tasks. We use batch size=32 and FP16 data type for all models. The inference speed and tuning time are shown in Figure~\ref{fig:e2e}. As we can see, \sys has significant better inference performance compared to Ansor. In particular, \sys is 4.2x faster on VGG models, 1.5x faster on ResNet models, and 2.6x faster on RepVGG models. On average, \sys improves the inference speed by 2.8x compared to Ansor. In terms of tuning time, as shown in Figure~\ref{fig:e2e_tuning}, \sys can complete the tuning much faster than Ansor because \sys uses hardware-native templated search which greatly reduces the searching space. Concretely, \sys can finish the tuning within 20 minutes for all models while Ansor takes 12 hours on average. 

\subsection{System-friendly models: RepVGG case study} 

Finally, we perform a case study on RepVGG~\cite{repvgg} to show the effectiveness of our system-model codesign principles. RepVGG is a VGG-like convolution neural network which only uses 3$\times$3 Conv2Ds to achieve higher accuracy and faster inference speed. The key idea of RepVGG is to train high-accuracy models with branches and remove those branches by re-parameterization to accelerate the inference. We apply our system-friendly model principles to augment RepVGG. In our experiments, models are trained on ImageNet with FP32  using the Swin Transformer training codebase~\cite{swin}, but they are quantized to FP16 for inference without accuracy loss. The inference runs on an NVIDIA Tesla T4 GPU using 32 as the batch size.

\textbf{Changing activation functions.} We first augment RepVGG by trying different activation functions. The original RepVGG model selects ReLU as its activation function, but we also experiment with GELU~\cite{gelu}, Hardswish~\cite{hardswish}, and Softplus~\cite{softplus}. The top-1 accuracy and inference speed of RepVGG-A0 with different activation functions is shown in Table~\ref{tab:activation}. We have found that activation functions do affect the accuracy---RepVGG-A0 with Hardswish achieves 0.67\% higher accuracy. Meanwhile, the inference speed does not show too much difference. Even with Softplus that has complex computation, the speed only drops by 7.7\%.

\begin{table}[!t]
\centering
\caption{The top-1 accuracy and speed of RepVGG-A0 using different activation functions (120 epochs + simple data augmentation).}
\vspace{1mm}
\scalebox{0.9}{
\begin{tabular}{p{20mm}||p{25mm}p{25mm}}
\Xhline{3\arrayrulewidth}
\hfil Activation & \multicolumn{1}{c}{\hfil Top-1 accuracy} & \multicolumn{1}{c}{\hfil Speed (images/sec)} \\ \Xhline{2\arrayrulewidth}
\hfil ReLU & \hfil 72.31 & \hfil 5909 \\ 
\hfil GELU & \hfil 72.38 & \hfil 5645 \\
\hfil Hardswish & \hfil 72.98 & \hfil 5713 \\ 
\hfil Softplus & \hfil 72.57 & \hfil 5453 \\ 
\Xhline{3\arrayrulewidth}
\end{tabular}
}
\vspace{-5mm}
\label{tab:activation}
\end{table}

\textbf{Deepening the model with $1\times$1 Conv2Ds.} We apply our 2nd codesign principle by adding $1\times1$ Conv2Ds after each $3\times3$ Conv2D (except for the last one which has too many output channels). The $1\times1$ Conv2Ds have the same input and output channels, with strides of (1, 1) and no padding. \sys will fuse adjacent $3\times3$ and $1\times1$ Conv2Ds using persistent kernels if the fusion is beneficial. To verify the effectiveness of each individual principle, we do not change the activation function in this experiment. As shown in Table~\ref{tab:add11}, adding $1\times1$ Conv2Ds can improve the accuracy with minimal speed loss. The accuracy is increased by 0.82\%, 0.77\%, and 0.74\% for RepVGGAug-A0, A1, and B0 respectively. Their speed drops by 15.3\% on average.

\begin{table}[!t]
\centering
\caption{The top-1 accuracy and speed of original RepVGG models and their augmentation with $1\times1$ Conv2Ds (200 epochs + simple data augmentation).}
\vspace{1mm}
\scalebox{0.9}{
\begin{tabular}{p{30mm}||ccc}
\Xhline{3\arrayrulewidth}
\hfil Model & Top-1 accuracy & Speed & Params \\ \Xhline{2\arrayrulewidth}
\hfil RepVGG-A0 & 73.05 & 7861 & 8.31 \\ 
\hfil RepVGG-A1 & 74.75 & 6253 & 12.79 \\ 
\hfil RepVGG-B0 & 75.28 & 4888 & 14.34 \\ \Xhline{2\arrayrulewidth}
\hfil RepVGGAug-A0 & 73.87 & 6716 & 13.35 \\ 
\hfil RepVGGAug-A1 & 75.52 & 5241 & 21.7 \\ 
\hfil RepVGGAug-B0 & 76.02 & 4145 & 24.85 \\
\Xhline{3\arrayrulewidth}
\end{tabular}
}
\vspace{-2mm}
\label{tab:add11}
\end{table}

\begin{table}[!t]
\centering
\vspace{-2mm}
\caption{The top-1 accuracy and speed of original RepVGG models and their augmentation with $1\times1$ Conv2Ds+Hardswish (300 epochs + advanced augmentation, label smoothing, and mixup).}
\vspace{1mm}
\scalebox{0.9}{
\begin{tabular}{p{25mm}||cc}
\Xhline{3\arrayrulewidth}
\hfil Model & Top-1 accuracy & Speed (images/sec) \\ \Xhline{2\arrayrulewidth}
\hfil RepVGG-A0 & 73.41 & 7861  \\ 
\hfil RepVGG-A1 & 74.89 & 6253  \\ 
\hfil RepVGG-B0 & 75.89 & 4888  \\ \Xhline{2\arrayrulewidth}
\hfil RepVGGAug-A0 & 74.54 & 6338 \\ 
\hfil RepVGGAug-A1 & 76.72 & 4868 \\ 
\hfil RepVGGAug-B0 & 77.22 & 3842 \\ 
\Xhline{3\arrayrulewidth}
\end{tabular}
}
\vspace{-4mm}
\label{tab:alltricks}
\end{table}


\textbf{Combined effect.} Finally, we combine the above two techniques and train the model with advanced augmentation, label smoothing, and mixup in 300 epochs. For RepVGG-A0, we train it for 300 epochs with only simple augmentation, which has better accuracy. 
As shown in Table~\ref{tab:alltricks}, designing models in a system-friendly manner can improve accuracy more efficiently. 
For example, in original RepVGG models, B0 is built by augmenting A1 with more $3\times3$ Conv2Ds, which has 1\% higher accuracy and 21.8\% lower speed compared to A1. In our augmentation, however, RepVGGAug-A1 is augmented by adding $1\times1$ Conv2Ds which has similar speed overhead, but the accuracy is improved by 1.83\% than RepVGG-A1. Note that designers have the flexibility to make a trade off between accuracy and speed. For instance, by adding only $1\times1$ Conv2Ds to the first three layers of RepVGG-A0 and using Hardswish, we can get a RepVGAug-A0 model with 74.02\% Top-1 accuracy and 7288 images/sec speed.

\section{Discussion}

\textbf{Other platforms.}
Although we use NVIDIA CUTLASS and GPUs to demonstrate the design of \sys, our approach is not bound to any specific devices or libraries. Templated libraries are a design trend for other devices as well. 
We expect that for other libraries, a \sys-like approach would also result in new end-to-end optimizations. 

\textbf{Persistent kernel fusion limitations.} Although persistent kernels can fuse any sequential GEMMs/Conv2Ds following the threadblock residence, we design it specifically for memory-bound operators, which is consistent with the motivation of general operator fusion. That is, \sys can improve the performance for sequential GEMMs with small N and K dimensions but large M dimensions and Conv2Ds with small channels. Fusing compute-bound operators could lead to performance drops because of the threadblock residence requirement. 

\section{Related Work}
\label{sec:related}

\textbf{Auto-tuners.} Many DNN compilers and frameworks employ auto-tuning strategies to search for the tensor implementation with optimal performance~\cite{autoTVM, ansor, halide-auto-scheduler, flextensor}. 
As they infer the hardware cost model by trying different tensor implementations and measuring their performance,  this takes hours to days. Moreover, the generated tensor programs cannot achieve hardware-native performance. 
\sys bridges the gap between auto-tuners and hardware-native performance.

\textbf{Operator fusion.} Operator fusion is an important graph-level optimization~\cite{tvm, tensorflow, xla, pytorch, learningtofuse, taso}. However, existing operator fusion only considers one GEMM/Conv and its adjacent operators, e.g., BiasAdd, ReLU, and the fusion is not well supported by vendor libraries. \sys enables new operator fusion with high performance. For instance, the proposed persistent kernel fusion can fuse a sequence of GEMMs and Convs, further improving the performance. Our persistent kernel is different from Persistent RNNs~\cite{diamos2016persistent} which is manually designed specifically for RNNs without using tensor cores.

\textbf{System-friendly model design.} RepVGG~\cite{repvgg} designs system-friendly models by employing multi-branch architectures in training models to pursue high accuracy and by removing the branches via re-parameterization for the inference. Also, RepVGG uses only $3\times3$ Conv2Ds which are well-supported by the hardware. \sys further extends the idea by proposing system-model codesign principles, and uses RepVGG as a concrete case study. 

\section{Conclusion}
\label{sec:conclusion} 

This paper presents \sys, which bridges the gap between auto-tuners and device library performance. \sys utilizes the emerging trend that vendor libraries are becoming modularized and composable. It combines the flexibility of auto-tuners and the hardware-native performance of templated device libraries to achieve the best of both worlds. Our design enables new tensor-level and graph-level optimizations, and inspires system-friendly model design insights. Our experiments show that \sys can achieve 2.5$\times$ speedup on widely-used convolutional neural networks compared against the state of the art. Moreover, it finishes its auto-tuning within 20 minutes.

\bibliography{paper}
\bibliographystyle{mlsys2021}

%


\end{document}